\documentclass[apj]{emulateapj}
\pdfoutput=1
\usepackage{hyperref}
\usepackage{amsmath,amstext}
\usepackage[T1]{fontenc}
\usepackage[figure,figure*]{hypcap}

\shorttitle{WD J2356$-$209}
\shortauthors{Blouin et al.}

\begin{document}

\submitted{Accepted for publication in The Astrophysical Journal}

\title{A New Generation of Cool White Dwarf Atmosphere Models.
  III. WD~J2356$-$209: Accretion of a Planetesimal with an Unusual Composition}

\author{S. Blouin\altaffilmark{1}}
\author{P. Dufour\altaffilmark{1,2}}
\author{N.F. Allard\altaffilmark{3,4}}
\author{S. Salim\altaffilmark{5}}
\author{R.M. Rich\altaffilmark{6}}
\author{L.V.E. Koopmans\altaffilmark{7}}

\altaffiltext{1}{D\'epartement de physique, Universit\'e de Montr\'eal, Montr\'eal, 
QC H3C 3J7, Canada; sblouin@astro.umontreal.ca, dufourpa@astro.umontreal.ca}
\altaffiltext{2}{Institut de recherche sur les exoplan\`etes,
  D\'epartement de physique, Universit\'e de Montr\'eal, Montr\'eal, 
  QC H3C 3J7, Canada}
\altaffiltext{3}{GEPI, Observatoire de Paris, Universit\'e PSL, CNRS, UMR 8111,
  61 avenue de l'Observatoire, F-75014 Paris, France}
\altaffiltext{4}{Sorbonne Universit\'e, CNRS, UMR 7095,
  Institut d'Astrophysique de Paris, 98bis boulevard Arago, F-75014 Paris, France}
\altaffiltext{5}{Department of Astronomy, Indiana University, Bloomington, IN
  47404, USA}
\altaffiltext{6}{Department of Physics and Astronomy, University of California,
  Los Angeles, CA 90095, USA}
\altaffiltext{7}{Kapteyn Astronomical Institute, University of Groningen,
  PO Box 800, NL-9700 AV Groningen, The Netherlands}

\begin{abstract}
  WD~J2356$-$209 is a cool metal-polluted white dwarf whose visible spectrum is
  dominated by a strong and broad sodium feature. Although discovered nearly
  two decades ago, no detailed and realistic analysis of this star had yet been
  realized. In the absence of atmosphere models taking into account the nonideal
  high-density effects arising at the photosphere of WD~J2356$-$209, the origin of
  its unique spectrum had remained nebulous. We use the cool white dwarf atmosphere
  code presented in the first paper of this series to finally reveal the secrets
  of this peculiar object and details about the planetesimal that polluted its atmosphere.
  Thanks to the improved input physics of our models,
  we find a solution that is in excellent agreement with the photometric observations
  and the visible spectrum. Our solution reveals that the photosphere of WD~J2356$-$209 
  has a number density ratio of $\log\,{\rm Na/Ca}= 1.0 \pm 0.2$, which is the highest
  ever found in a white dwarf. Since we do not know how long ago the accretion
  episode stopped (if it has), we cannot precisely determine the composition nor the mass
  of the accreted planetesimal. Nevertheless, all scenarios considered indicate that
  its composition is incompatible with that of chondrite-like material and that its
  mass was at least $10^{21}\,{\rm g}$.
\end{abstract}

\section{Introduction}
WD~J2356$-$209\footnote{Also known as WD~2354$-$211 in the electronic version of
of the Catalog of Spectroscopically Identified White Dwarfs \citep{mccook1999catalog}},
was discovered by \cite{oppenheimer2001direct} as part of their search
for cool white dwarfs in the galactic halo. Since its spectrum shows a broad asymmetric
feature around 6000{\,\AA}, it was described as having a "bizarre spectrum, 
incomparable to any other known object".

Using only $BRI$ photometry, \cite{bergeron2003critical} attempted the first atmospheric
parameter determination of WD~J2356$-$209. It was found that it must have an effective
temperature of the order of $4000\,{\rm K}$: the best photometric solution was at
$T_{\rm eff}=3400\,{\rm K}$ if a hydrogen-rich atmosphere was assumed and 
at $T_{\rm eff}=4610\,{\rm K}$ in the case of a helium-rich composition.
Note that because of the absence of Balmer lines at such low temperatures, the atmospheric
composition of this object remained unknown.

Shortly after, \cite{salim2004cool} reobserved WD J2356$-$209 and confirmed the measurements
of \cite{oppenheimer2001direct}. In particular, in a color-color diagram,
it appears as a strong outlier 
with an excessively blue $B-V$ and an extremely red $V-I$. \cite{salim2004cool} suggested
that these peculiar colors may be the result of a very broad \ion{Na}{1} D doublet, implying that
WD~J2356$-$209 is a metal-polluted white dwarf (i.e., a DZ star).

Then, \cite{bergeron2005interpretation} obtained new photometric observations of
WD~J2356$-$209 in the $BVRI$ and $JH$ bands. They picked up the interpretation proposed
by \cite{salim2004cool} and analyzed WD~J2356$-$209 with atmosphere models that included Na.
Their solution ($T_{\rm eff}=4790\,{\rm K}$ and $\log\,{\rm Na/He}=-5$) 
yields a satisfactory fit both to their photometric measurements and
to the spectrum of \cite{oppenheimer2001direct} in the 5000--9000{\,\AA} region. 
However, in the absence of any spectrum below 5000{\,\AA} they could not detect any
other metal than Na, and thus Na and He were the only atomic species included in their
atmosphere model. In retrospect, this was an unrealistic assumption.
With the discovery of dozens of circumstellar debris discs around metal-polluted white dwarfs 
\citep[e.g.,][]{zuckerman1987excess,gansicke2006gaseous,farihi2009infrared,melis2010echoes,rocchetto2015frequency}
and the detection of planetary transits in the light curve of WD~1145+017
\citep{vanderburg2015disintegrating,gansicke2016high,croll2017multiwavelength}
it is now clear that the presence of heavy elements in white dwarfs is the
consequence of the accretion of tidally disrupted rocky bodies 
\citep{jura2003tidally,farihi2010rocky,jura2014extrasolar}.
Therefore, all elements representative of the composition of rocky planetesimals
should be included in DZ atmosphere models. 

Moreover, the analysis of \cite{bergeron2005interpretation}
was based on models that did not take into account the nonideal high-density effects that are known to
arise under the physical conditions met in the atmospheres of cool white dwarfs. In particular,
their models relied on the ideal gas law, which is inappropriate for the high pressures of cool white dwarfs
\citep[e.g.,][]{saumon1995equation,becker2014ab};
nonideal effects affecting the chemical equilibrium \citep{kowalski2007equation,blouin2018model} were neglected;
and simple Lorentzian line profiles, which poorly reproduce the spectral lines observed in cool DZ
white dwarfs \citep{allard2016asymmetry,allard2018line,hollands2017cool}, were assumed.

An independent analysis of WD~J2356$-$209 was performed by \cite{homeier2005,homeier2007}
using the PHOENIX general-purpose stellar atmosphere code \citep{phoenix1,phoenix2,phoenix3}
and the spectrum obtained by \cite{oppenheimer2001direct}.
The first results identified an apparent overabundance of Na relative to Ca and suggested
that WD~J2356$-$209 has a hydrogen-rich atmosphere \citep{homeier2005}. An important limitation
of this first analysis was that line profiles implemented in their atmosphere models
were limited to perturber densities not exceeding $10^{19}\,{\rm cm}^{-3}$ 
\citep{allard2003new}, which is about three orders of magnitude below the density
at the photosphere
of a $T_{\rm eff}=4000\,{\rm K}$ helium-rich DZ white dwarf \citep{blouin2018model}.
In a second analysis
\citep{homeier2007}, improved line profiles appropriate for larger densities were employed and
a helium-rich composition was favored. \cite{homeier2005,homeier2007} were also
the first to point out that metal hydrides (namely, MgH and CaH) should form under the 
conditions encountered in the atmosphere of WD~J2356$-$209. However, 
these preliminary analyses were exploratory in nature and no quantitative fit to
the spectroscopic and photometric data was attempted. Furthermore, as in 
\cite{bergeron2005interpretation}, their models did not include the nonideal effects that
are known to affect the chemical equilibrium and the equation of state.

In this paper, the third of a series, we present the most detailed analysis to
date of WD~J2356$-$209. Our models, which
include an accurate description of the high-pressure physics relevant to the
modeling of cool DZ stars (see \citealt{blouin2018model,blouin2018dzcia}, hereafter
Papers~I and II),
make it possible for the first time to obtain a
satisfactory fit to the photometric data and to the entire visible spectrum.
In Section \ref{sec:observations}, we present the observations on which
our analysis of WD~J2356$-$209 is based.
A few new improvements brought to our model atmosphere code
are detailed in Section \ref{sec:models}. Section \ref{sec:analysis} contains
our analysis of WD~J2356$-$209, where we find that the sodium to calcium ratio is the
highest ever encountered in a white dwarf photosphere.
Finally, Section \ref{sec:interpretation} provides a discussion on the origins of the peculiar
abundances measured in WD~J2356$-$209 and our conclusions are given in Section \ref{sec:conclusion}.

\section{Observations}
\label{sec:observations}
Our analysis of WD~J2356$-$209 makes use of $BVRI$ and $JH$ photometry from
\cite{bergeron2005interpretation}, $grizy$ photometry from Pan-STARRS
\citep{chambers2016panstarrs} and the \textit{Gaia} DR2 parallax measurement
\citep{prusti2016gaia,brown2018gaia}. The parallax indicates that
WD~J2356$-$209 is located at a distance $D = (64.8 \pm 2.5)\,{\rm pc}$ from
the Earth, which is in good agreement with the estimation obtained by
\cite{salim2004cool} using color-magnitude relations ($74 \pm 34\,{\rm pc}$).

\begin{figure}
    \includegraphics[width=\columnwidth]{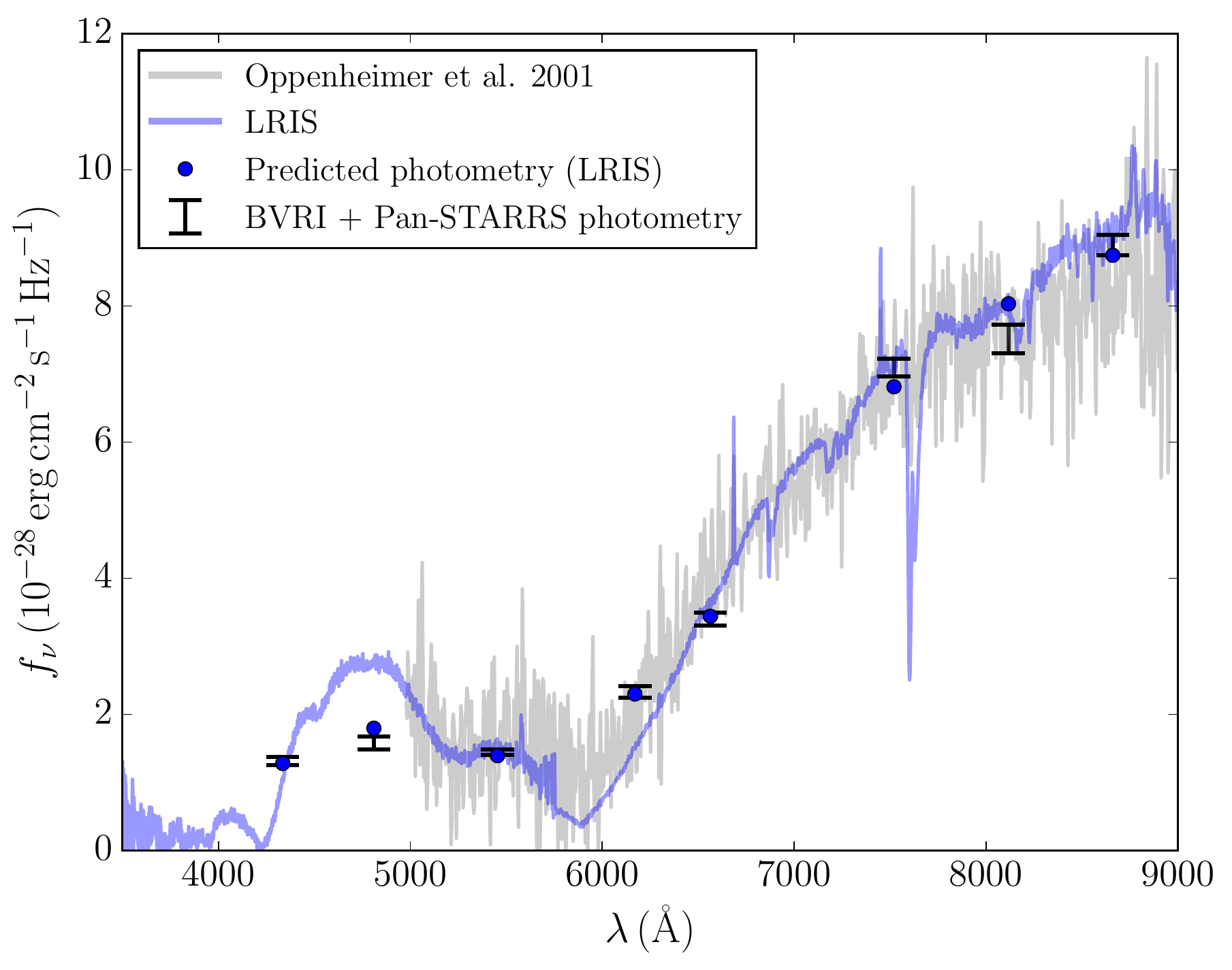}
    \caption{Observed spectra and photometry of WD~J2356$-$209. The spectrum of
      \cite{oppenheimer2001direct} is in gray and the Keck LRIS
      spectrum is in blue. Note that the LRIS spectrum was scaled so
      that the predicted photometry (blue circles) match the observed
      photometry (black error bars).}
  \label{fig:2356obs}
\end{figure}

For the spectroscopy, we rely on data obtained
with LRIS \citep{oke1995keck} on Keck I
telescope on 2002 September 14.
A 300/5000 grating was used for both the
blue and the red spectra, together with a 1 arcsec slit, giving an
effective resolution of 10.3\,{\AA}. D560 dichroic was used to split the
beams, giving useful wavelength coverage of 5000$-$9000\,{\AA} on the red
side and 3500$-$5800\,{\AA} on the blue side. Each of the blue and red
spectra were obtained with a total exposure time of 45 minutes. Standard
arc spectra were taken for wavelength calibration. Only one standard
star was available for flux calibration, while there were no suitable
observations for the removal of the telluric absorption. The
observations of the standard star may have been affected by detector
non-linearity. Therefore, we correct the flux calibration by scaling the
spectrum so that the photometry extracted from the joined blue/red
spectrum agrees with the broad-band photometry.
As shown in Figure \ref{fig:2356obs}, the scaled LRIS spectrum
agrees very well with the low-resolution shallow spectrum of
\cite{oppenheimer2001direct} in the region where they overlap ($>5000\,${\AA}).

\section{Atmosphere models}
\label{sec:models}
The atmosphere models used in this work are described at length in Paper I.
Our models are uniquely suited for the analysis of such a very cool DZ star, since they are
the only ones to include both a state-of-the-art description of the
ionization equilibrium of heavy elements (Paper I) and unified line profiles
  \citep{allard1999effect} for the strong Na and Ca spectral lines that characterize
  the spectrum of WD~J2356$-$209.

Compared to the models used in Papers I and II, two changes were made to our code. 
First, we have improved our collisional profiles of the resonance 
line of neutral calcium perturbed by helium using the ab initio Ca$-$He potential energies of 
M. Kro{\'s}nicki reported in \cite{hernando2008absorption}. It is a significant improvement in 
the description of Ca$-$He singlet potential energies, but only at close range
(i.e., only for radiator-perturber separations of $\approx 3.5$ to 11{\,\AA}).
To predict the impact core shift and width, the long-range part of the potential energies 
needs to be accurately determined. Moreover, our approach requires prior knowledge of the 
variation of the radiative dipole moments with atom-atom separation for each molecular state.
As this variation is unknown at present, the dipole moments were assumed to remain constant 
throughout the collision. In order to obtain more 
accurate line profiles, intensive ab initio calculations are being performed to obtain both 
the ground and first excited potential energy curves (PECs) and the transition dipole 
moments for the Ca$-$He system (N.~F. Allard et al., in preparation).

The second change made to our atmosphere model code was to add the opacity due
to the rovibrational transitions of the MgH and CaH molecules.
The MgH opacities are computed using the linelists of \cite{GharibNezhad2013einstein}
for the the X$^2\Sigma^+ \rightarrow$ A$^2\Pi$ and X$^2\Sigma^+ \rightarrow$ B$'^2\Sigma^+$
transitions, which are available on the ExoMol website\footnote{\url{http://exomol.com}}
\citep{tennyson2012exomol,tennyson2016exomol}.
For CaH, the X$^2\Sigma^+ \rightarrow$ A$^2\Pi$ and X$^2\Sigma^+ \rightarrow$ B$^2\Sigma^+$
transitions are computed
using the Kurucz linelists\footnote{\url{http://kurucz.harvard.edu}}, which
rely on molecular data from \cite{weck2003vibration} and \cite{shayesteh2013fourier}.

\section{Analysis of WD~J2356$-$209}
\label{sec:analysis}
We rely on the $BVRIJH$ and $grizy$ photometry of WD~J2356$-$209 to find its effective
temperature, surface gravity and H/He abundance ratio. More specifically, the solid angle
$\pi \left( R/D \right)^2$ and $T_{\rm eff}$ are found by fitting the model fluxes to the
spectral energy distribution. Since the distance $D$ is already known from the \textit{Gaia} parallax, we
can obtain the radius $R$ from the solid angle. The mass of the white dwarf (and the corresponding 
surface gravity) are then found using the evolutionary models of \cite{fontaine2001potential}
with C/O cores, $\log \left( M_{\rm He} / M_{\star} \right) = -2$ and
$\log \left( M_{\rm H} / M_{\star} \right) = -10$, which are representative of
helium-atmosphere white dwarfs.
The H/He abundance ratio is inferred from the infrared photometric measurements, which
are affected by the presence of H$_2-$He collision-induced absorption 
\citep[CIA,][]{jorgensen2000atmospheres,blouin2017cia}.
Once a photometric solution is found, we use
the spectroscopic data to constrain the abundances of Na, Ca, Fe and Mg.
In particular, the \ion{Na}{1} D doublet is used to derive the Na/He ratio, the \ion{Ca}{1} 4226{\,\AA}
and \ion{Ca}{2} H \& K lines for Ca/He, \ion{Fe}{1} and \ion{Fe}{2} lines in the 3600--3800{\,\AA}
region for Fe/He, and the MgH rovibrational bands between 5000 and 5300{\,\AA}
for Mg/He. Once the heavy element abundance ratios are found, the whole fitting procedure -- 
including the photometric fit -- is repeated until internal consistency is reached.

\begin{figure}
    \includegraphics[width=\columnwidth]{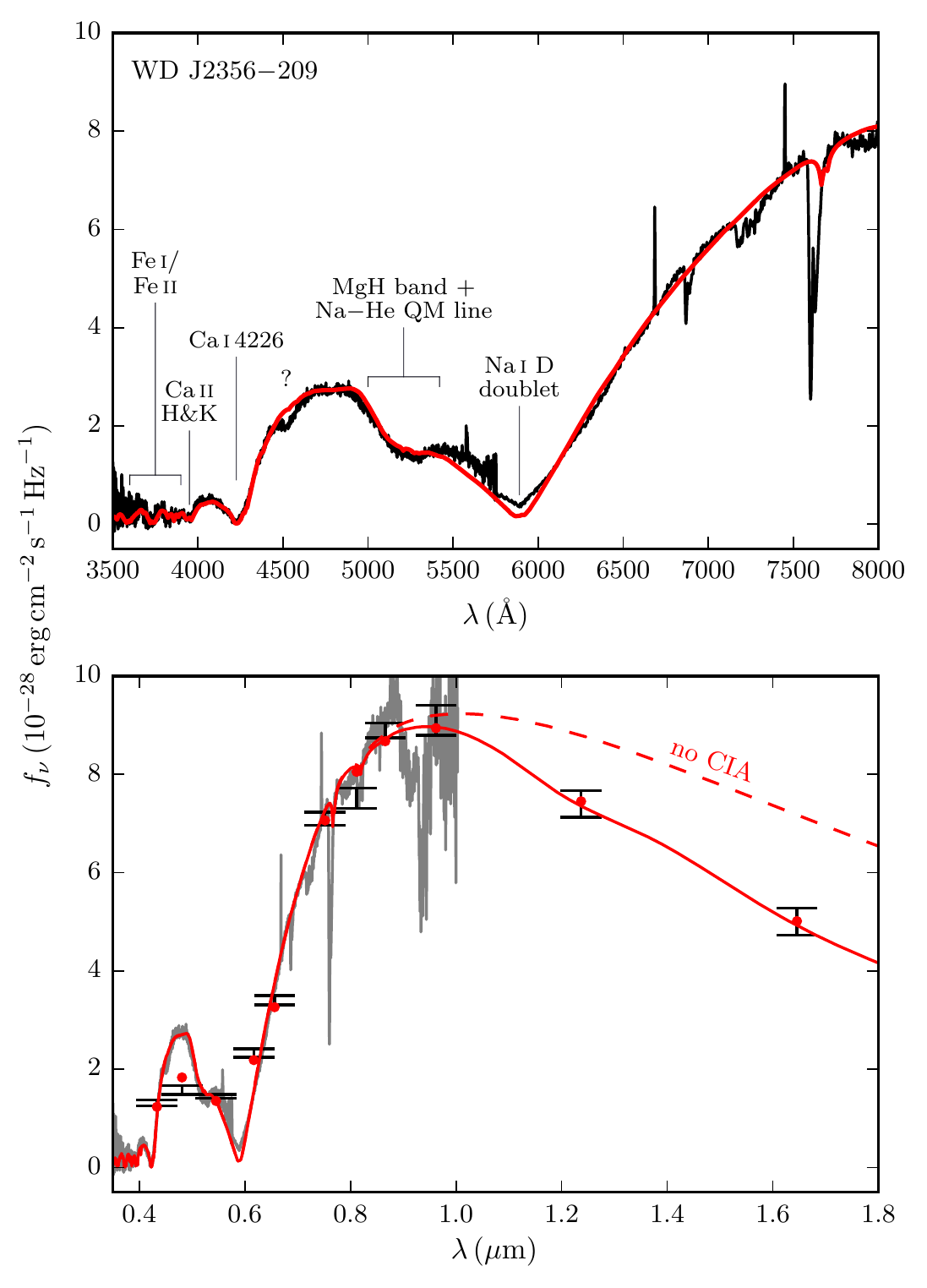}
    \caption{Our best solution for WD~J2356$-$209. The top panel shows our fit (in red)
    to the visible spectrum (in black) and the bottom panel shows our photometric
    fit to the $BVRI$, $JH$ and $grizy$ bands (the photometric observations are represented
    by the error bars). The dashed spectrum in the bottom panel
    was computed from our best-fitting model, but
    without including H$_2-$He CIA in the synthetic spectrum calculation.
    The strong absorption lines near 7600{\,\AA} are of telluric origin and were ignored
    in our analysis.}
  \label{fig:2356fit}
\end{figure}

Figure \ref{fig:2356fit} compares our best solution to the photometric and spectroscopic data.
Except for the core of the \ion{Na}{1} D doublet and an unknown feature near 4500{\,\AA}, 
our fit is in excellent agreement with the observations across all wavelengths. In particular,
our spectroscopic fit closely matches the Fe lines at small wavelengths, the
\ion{Ca}{2} H \& K and \ion{Ca}{1} 4226{\,\AA} lines, the blue and red wings of the
\ion{Na}{1} D doublet, and the flux depletion between 5000 and 5500{\,\AA}.
Note that due to strong telluric absorption we could not use the \ion{Na}{1}
  doublet at 8200{\,\AA} to validate the sodium abundance inferred from the \ion{Na}{1} D doublet.
Regarding the photometric fit, we found that a mix H/He atmosphere has to be assumed to produce the
H$_2-$He CIA that is required to properly match the $J$ and $H$ bands. 

The atmospheric parameters of WD~J2356$-$209 are given in Table \ref{tab:2356}.
Its very cool temperature of $T_{\rm eff}=4040 \pm 110\,{\rm K}$ makes
it -- to our knowledge -- the oldest known DZ star, with a cooling age of
$8.0 \pm 0.8\,{\rm Gyr}$.\footnote{Note that \cite{giammichele2012know}
    found a greater cooling age ($8.5\,{\rm Gyr}$) for WD~2251$-$070.
    However, their analysis
    was based on a model atmosphere code \citep{dufour2007spectral} that did not include a detailed
    description of the high-density effects arising at the photosphere of such a cool
    object. Our reanalysis of WD~2251$-$070 (Section \ref{sec:4500}) suggests
    that it has an effective temperature $T_{\rm eff}=4050 \pm 60\,{\rm K}$ and
    a surface gravity $\log g=7.94 \pm 0.04$, which corresponds to a cooling age
    of $7.5 \pm 0.5\,{\rm Gyr}$.}
  
Table \ref{tab:2356} also lists the constraints on metal abundance
ratios in the atmosphere of WD~J2356$-$209. Apart from the sodium abundance,
which is further discussed in Section \ref{sec:interpretation}, all abundances
are close to the average values measured in metal-polluted atmospheres.
In particular, in a ternary diagram of Ca, Mg and Fe abundances
\citep[e.g.,][Figure 2]{hollands2018cool}, WD~J2356$-$209 would be close to the bulk of DZ stars.
The constraints on Al, K, Ti, Cr, Mn and Ni,
for which no feature is observed in the spectrum of WD~J2356$-$209,
were found by raising their abundances up to the point where a spectral
line should be visible. Note that the constraint on the K/He ratio was found using
the spectrum of \cite{oppenheimer2001direct} instead of the LRIS spectrum, since the
latter shows strong telluric features near the \ion{K}{1} 7665/7699{\,\AA} lines
(see Figure \ref{fig:2356obs}). The constraint on C/He was established using the C$_2$
rovibrational bands instead of the atomic carbon lines, since the former require a smaller
carbon abundance before being visible in the spectrum. Finally, no firm constraints could be
found for the abundances of O and Si, since even the addition of a very large amount of
these elements does not result in any visible spectral lines. The limits for O/He and Si/He
given in Table \ref{tab:2356} are therefore lower-bound estimates of the maximum
abundance of these elements that is compatible with the spectroscopic observations of
WD~J2356$-$209.

\begin{deluxetable}{cc}
  \tablecaption{WD~J2356$-$209 atmospheric parameters \label{tab:2356}}
  \tablehead{\colhead{\hspace{0.7cm}Parameter}\hspace{0.7cm} &
    \colhead{\hspace{0.7cm}Value}\hspace{0.7cm}}
  \startdata
  $T_{\rm eff}$       & $\phn \phn \phd 4040 \pm 110\,{\rm K}$\\
  $\log g$            & $\phn \phd 7.98 \pm 0.07$\\
  $\log\,{\rm H/He}$\tablenotemark{a}  & $-1.5 \pm 0.2$\\
  $\log\,{\rm C/He}$  & $<-6.0$ \\
  $\log\,{\rm O/He}$  & $(<-4)$\tablenotemark{b}\phd\\
  $\log\,{\rm Na/He}$ & $-8.3 \pm 0.2$\\
  $\log\,{\rm Mg/He}$ & $-8.0 \pm 0.2$\\
  $\log\,{\rm Al/He}$ & $<-8.8$\\
  $\log\,{\rm Si/He}$ & $(<-6.8)$\\
  $\log\,{\rm K/He}$  & $<-10.0$\\
  $\log\,{\rm Ca/He}$ & $-9.3 \pm 0.1$\\
  $\log\,{\rm Ti/He}$ & $<-9.8$\\
  $\log\,{\rm Cr/He}$ & $<-9.9$\\
  $\log\,{\rm Mn/He}$ & $<-10.7$\\
  $\log\,{\rm Fe/He}$ & $-8.6 \pm 0.2$\\
  $\log\,{\rm Ni/He}$ & $<-8.0$
  \enddata
  \tablenotetext{a}{All abundances are reported as ratios of number densities}
  \tablenotetext{b}{The abundances in parentheses are not firm limits (see text)}
\end{deluxetable}

Our excellent fit to the wings of the broad Ca and Na spectral lines of WD~J2356$-$209 (Figure \ref{fig:2356fit}) 
was made possible
thanks to the improved lines profiles implemented in our models \citep{allard2014caii,allard2014nai}.
This good agreement
indicates that our model predicts the right physical conditions in the line-forming regions of the
atmosphere. In particular, it suggests that the total number density $n_{\rm tot}$ is accurate, 
since it is the parameter that governs the broadening of the spectral lines. 
At the photosphere of WD~J2356$-$209, the pressure is high enough ($\log P = 10.0$) that deviations 
from the ideal gas law begin to be important (see Figure 4 of Paper I). 
If the ideal gas law was assumed, $n_{\rm tot}$ would be 20\% higher and the line profiles
would be slightly broader (particularly the \ion{Na}{1} D doublet).
Note though that WD~J2356$-$209 cannot be used to validate our implementation of pressure
ionization, since its photospheric density ($\rho = 0.1\,{\rm g}\,{\rm cm}^{-3}$) is too small
to induce significant deviations from the ideal Saha equation \citep[][Paper I]{kowalski2007equation}.

\subsection{Collision-induced absorption}
\begin{figure}
    \includegraphics[width=\columnwidth]{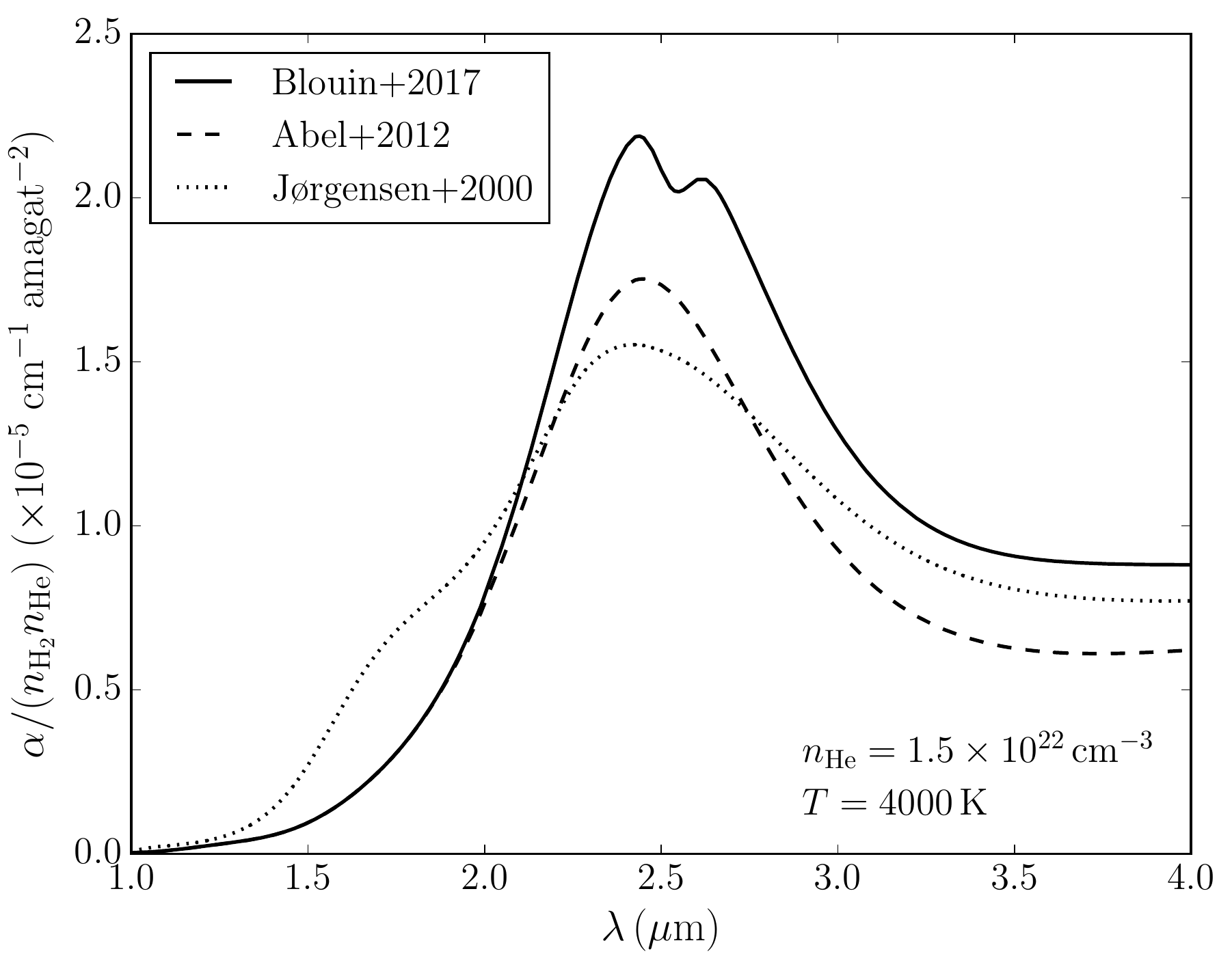}
    \caption{H$_2-$He CIA profiles at $4000\,{\rm K}$ and $n_{\rm He}=1.5 \times 10^{22}\,{\rm cm}^{-3}$,
      as computed by \citet[Equation 7]{blouin2017cia}, \cite{abel2012infrared} and \cite{jorgensen2000atmospheres}. 
      The absorption profiles are divided by the number density of H$_2$ and He and are therefore 
      constant with respect to density (with the exception of the \citeauthor{blouin2017cia} profile, which
      includes the effects of many-body collisions).}
  \label{fig:cia}
\end{figure}

Figure \ref{fig:cia} compares the H$_2-$He CIA profiles obtained by different authors
for physical conditions that are representative of the photosphere of WD~J2356$-$209. Differences
between the profiles imply that the H/He ratio derived from the photometry depends on the choice of 
the CIA profiles implemented in our models. In particular, if the
profiles of \cite{jorgensen2000atmospheres} were assumed, a smaller hydrogen abundance
would be found. In fact, as already noted
in Paper II, the profiles of \cite{jorgensen2000atmospheres} predict a too strong
absorption in the $\approx 1.2-2\,\mu{\rm m}$ region, possibly because the potential energy and induced
dipole surfaces used to derive those profiles were computed with a smaller atomic orbital basis set. 

Furthermore, Figure \ref{fig:cia} shows that significant differences between the profiles
of \cite{abel2012infrared} and \cite{blouin2017cia} appear above $2\,\mu{\rm m}$ (particularly
near the maximum of the fundamental band at $\approx 2.5\,\mu{\rm m}$). Those differences
are due to many-body collisions (which are only included in the profiles of \citeauthor{blouin2017cia})
that lead to an enhancement and a distortion of the absorption profile. The only photometric measurements
available beyond the $H$ band are those from the Wide-field Infrared Survey Explorer
\citep[WISE,][]{wright2010wide} for the 3.4 and 4.6\,$\mu$m bands. Our best fit is compatible with those
measurements, but their large uncertainties do not allow us to explicitly confirm that 
the CIA is indeed enhanced by many-body effects.

\subsection{The Na$-$He satellite and the MgH bands}
\begin{figure}
    \includegraphics[width=\columnwidth]{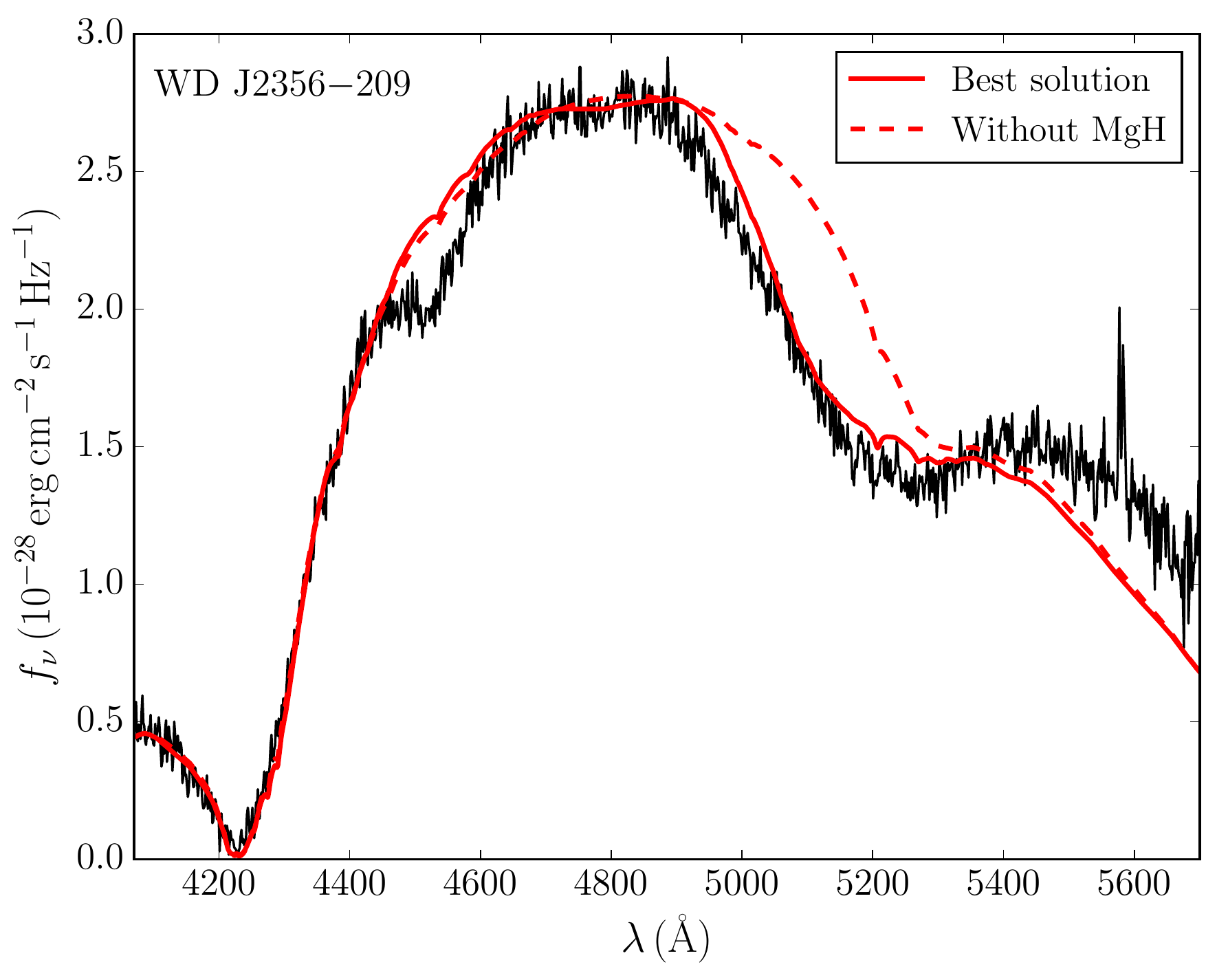}
    \caption{Comparison between the best solutions found when including (solid line) and when
    omitting (dashed line) the MgH bands in our models.}
  \label{fig:2356mgh}
\end{figure}

The wide absorption feature in the blue wing of the \ion{Na}{1} D doublet proved particularly
challenging to model. \cite{allard2014nai} suggested that it could be the result of the
quasi-molecular satellite feature arising from the Na$-$He interaction. Our detailed analysis
reveals that it is only part of the explanation and that MgH rovibrational bands must be included
to fully explain this absorption feature. Figure \ref{fig:2356mgh} compares our best fit to the
5000--5500{\,\AA} absorption feature with and without the MgH bands. Clearly, the Na$-$He satellite
visible in the fit that omits the MgH bands (dashed line) is not sufficient to explain the whole
absorption feature and MgH absorption bands must be invoked. 

We emphasize that including MgH in our models is physically motivated.
Given the temperature of WD~J2356$-$209, its hydrogen
abundance (independently constrained from the H$_2-$He CIA) and the presence of magnesium
(which is given by a Mg/Ca ratio that is very close to the value expected for chondrite-type material),
MgH has to be present at the photosphere of WD~J2356$-$209. To our knowledge,
this is the first detection of MgH molecular features in a white dwarf. MgH bands
were proposed by \cite{dufour2006model} to explain the asymmetry of the \ion{Mg}{1} 5175{\,\AA} line
in G165$-$7, but it was later shown that improved line profiles -- beyond the impact
approximation -- are enough to explain this asymmetry \citep{allard2016asymmetry}.
Finally, note that for our derived atmospheric parameters CaH is not sufficiently abundant to produce
visible bands around 7000{\,\AA}.

\subsection{The unknown absorption feature near 4500\,\AA}
\label{sec:4500}
Our fit of WD~J2356$-$209 does not reproduce the small asymmetric absorption feature
near 4500{\,\AA} (Figure \ref{fig:2356mgh}). Technically, boosting the Ti/Ca ratio
by a factor of $\approx 10$ with respect to chondritic abundances allows the
\ion{Ti}{1} $a^5F - y^5F^{\circ}$ multiplet (centered
at 4535{\,\AA}) to be strong enough to reasonably match the absorption feature 
near 4500{\,\AA}. However, another cool DZ star, WD~2251$-$070, offers good reasons to
be suspicious of this interpretation.

\begin{figure}
    \includegraphics[width=\columnwidth]{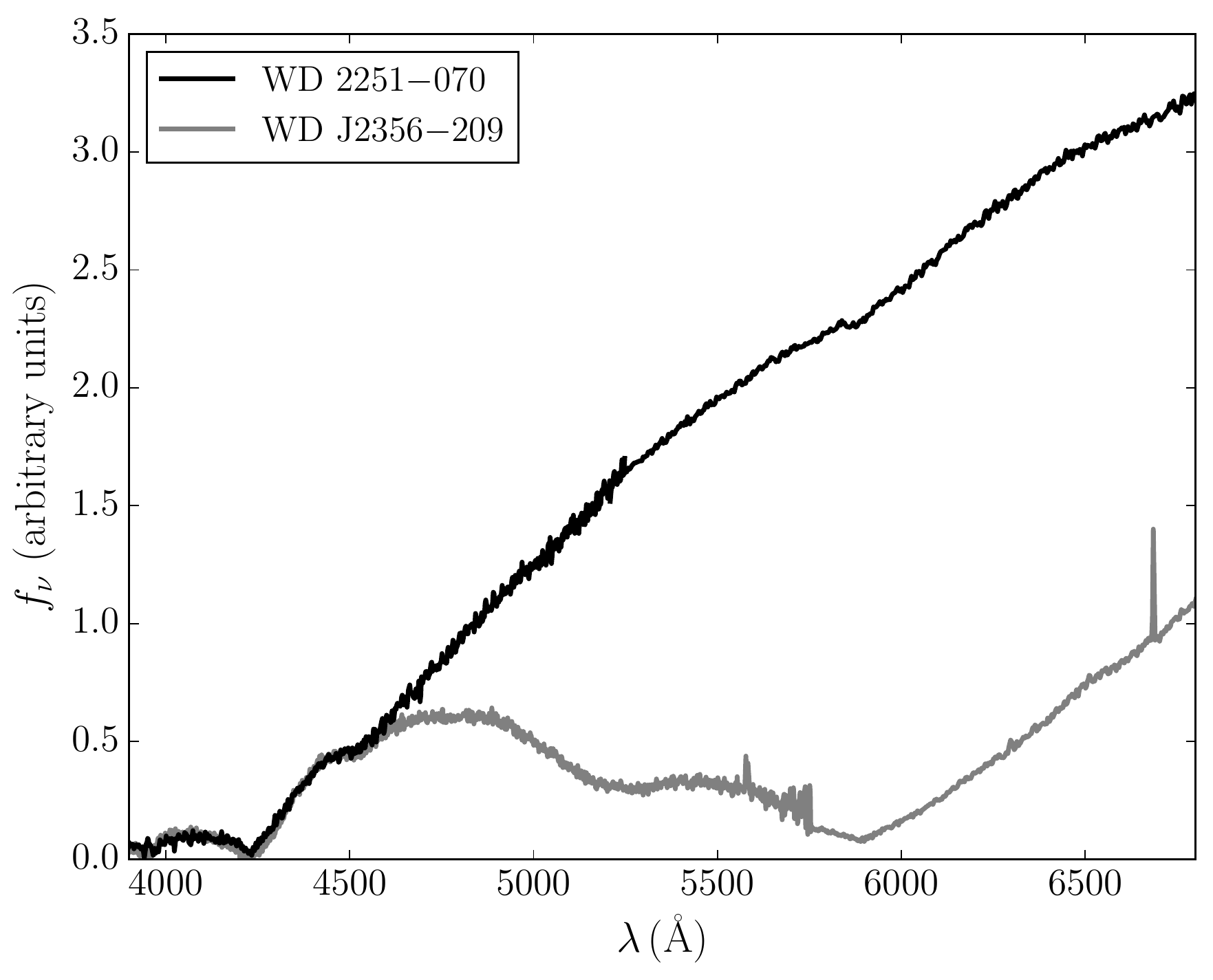}
    \caption{Comparison of the spectrum of WD~J2356$-$209 to that of
      WD~2251$-$070. The data for WD~2251$-$070 is from
      \cite{bergeron1997chemical} and \cite{dufour2007spectral}.
      Note the resemblance between both spectra
      below 4600{\,\AA} and their strong dissimilarity at longer wavelengths.
    }
    \label{fig:2251comparaison}
\end{figure}

Indeed, WD~2251$-$070 also shows a strikingly similar absorption feature in that region
(Figure \ref{fig:2251comparaison}). Using photometric and spectroscopic data reported in
\cite{bergeron1997chemical} and \cite{dufour2007spectral}, we analyzed WD~2251$-$070 with
our cool DZ grid. Our preliminary fit yields $T_{\rm eff}=4050 \pm 60\,{\rm K}$,
$\log g = 7.94 \pm 0.04$, $\log\,{\rm H/He} <-4.5$, $\log\,{\rm Na/He} < -10.3$
and $\log\,{\rm Ca/He} = -10.0 \pm 0.1$ (an in-depth analysis of this object will
be presented in N.~F. Allard et al., in preparation). 
Given the strong similarities between WD~2251$-$070 and WD~J2356$-$209 (i.e., same feature near 4500{\,\AA},
similar Ca abundances and virtually identical effective temperatures), 
it is very likely that the origins of the 
4500{\,\AA} absorption feature are the same for both objects.
If it is the case, then there is a good reason to reject Ti as the explanation for this feature.
Boosting the Ti ratio to produce a sufficiently strong 
\ion{Ti}{1} $a^5F - y^5F^{\circ}$ multiplet also implies that
the similarly important $a^5F - y^5 G^{\circ}$ multiplet (centered around 5020{\,\AA})
should be visible. By adjusting the Mg abundance, our fit of WD~J2356$-$209 could accommodate
the additional opacity resulting from the \ion{Ti}{1} $a^5F - y^5 G^{\circ}$
multiplet. However, the spectrum of 
WD~2251$-$070 clearly rules out the presence of any absorption feature around 5020{\,\AA}
and thus suggests that Ti cannot be the explanation of the 4500{\,\AA} absorption feature.

Still, since we rely on Lorentzian line profiles for
all Ti spectral lines, we need to be careful before completely ruling out Ti as the source of
the 4500{\,\AA} feature.
In principle, it is possible that accurate line profile calculations will show that the
$a^5F - y^5 G^{\circ}$ multiplet is flattened out under the density
conditions encountered at the photosphere of WD~2251$-$070 ($n_{\mathrm He} \approx 6 \times 10^{22}\,{\rm cm}^{-3}$).
However, there is a second argument against the Ti scenario. We constrained the Na abundance in
WD~2251$-$070 to $\log\,{\rm Na/He}<-10.3$, implying that it has a $\log\,{\rm Na/Ca}$ ratio smaller than $-0.3$.
This ratio is vastly different from the $\log\,{\rm Na/Ca}=1.0$ ratio found for WD~J2356$-$209 and we therefore
expect the abundance pattern of WD~2251$-$070 to be quite different from that of WD~J2356$-$209. In particular,
it would be surprising if WD~2251$-$070 and WD~J2356$-$209 had similar Ti/Ca ratios but 
significantly different Na/Ca ratios. All things considered, it is unlikely that the feature
near 4500{\,\AA} is due to Ti.

Alternatively, since it is observed in two white dwarfs with similar Ca abundances,
it is tempting to explain the 4500{\,\AA} feature as being due to Ca.
In a dense helium medium, it turns out that the \ion{Ca}{2} H \& K profile 
can show a quasi-molecular feature near 4500{\,\AA} \citep{allard2014caii}. However, for the physical
conditions in the atmosphere of WD~2251$-$070 and WD~J2356$-$209, this quasi-molecular line is
predicted to be too weak to explain the shape of the spectrum around 4500{\,\AA}. We are therefore unable
to identify the origin of this feature at the moment.

\section{How to explain the high sodium abundance?}
\label{sec:interpretation}
As it is well above the chondritic value of $\log\,{\rm Na/Ca}=0.0$ \citep{lodders2003solar},
the $\log\,{\rm Na/Ca}=1.0\pm0.2$ ratio measured in WD~J2356$-$209 is surprising and needs
to be explained. Note that this extreme abundance ratio is not only supported by our excellent fit
to the observations (Figure \ref{fig:2356fit}), but also by the direct comparison of 
WD~J2356$-$209 to WD~2251$-$070. While both objects have a very similar calcium abundance and 
a virtually identical effective temperature, their spectra are drastically different in the 
region affected by the \ion{Na}{1} D doublet (as seen in Figure \ref{fig:2251comparaison}).

\subsection{Comparison to other sodium-rich stars}

\begin{figure}
    \includegraphics[width=\columnwidth]{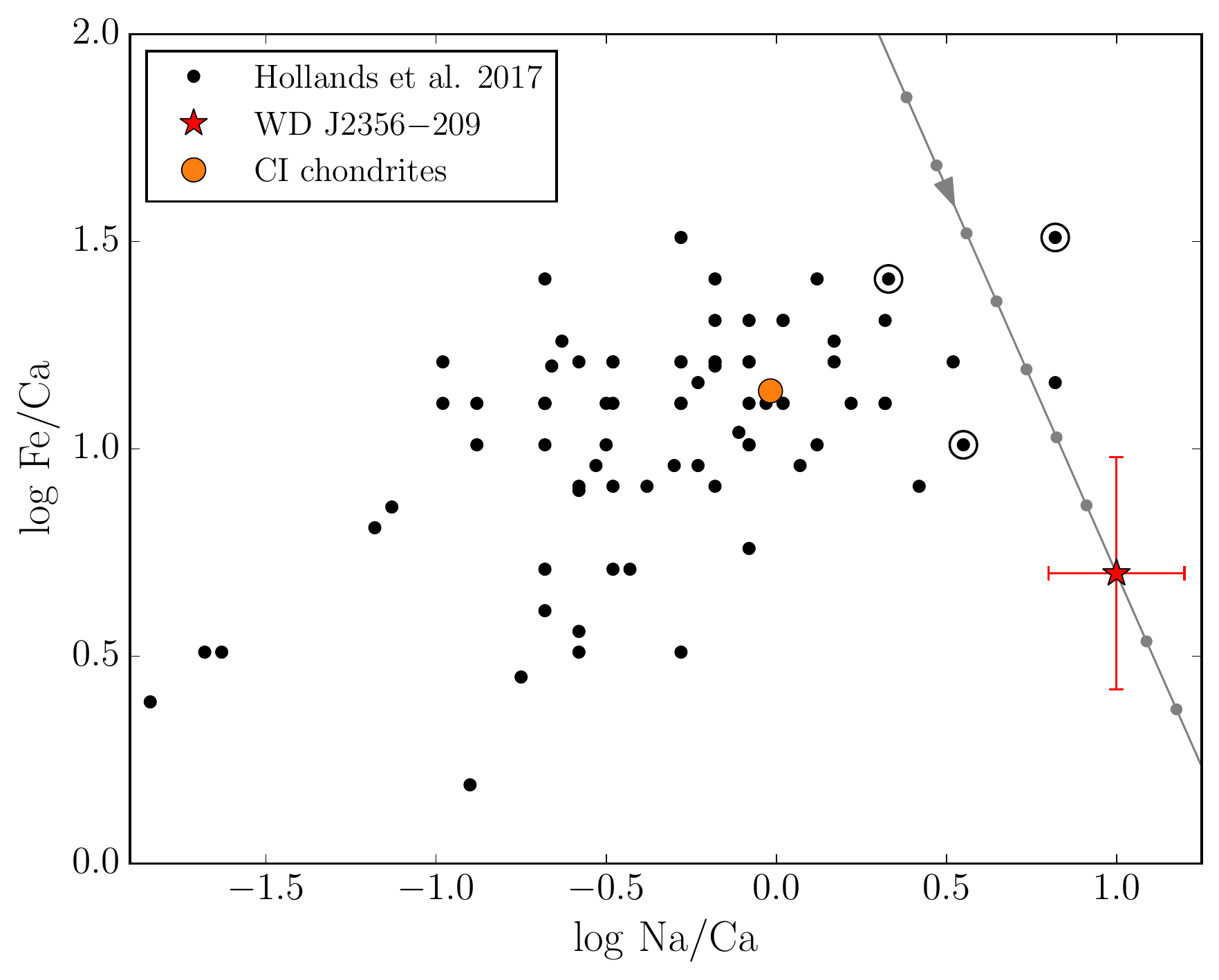}
    \caption{Elemental abundances for CI chondrites
      \citep[][filled orange circle]{lodders2003solar}, WD~J2356$-$209 (red star)
      and the DZ stars analyzed by \citet[filled black circles]{hollands2017cool}.
      The three encircled objects are those for which we found compelling evidence
      of a Na/Ca abundance ratio greater than 1.
      Note that magnetic objects in the \cite{hollands2017cool} sample
      are not shown in this figure.
      The gray line indicates how the metal abundances in WD~J2356$-$209
      would have evolved, assuming the diffusion timescales mentioned in the text.
      The distance between two gray circles on the chemical evolution track corresponds
      to 1~Myr.}
  \label{fig:diffusion_evolution}
\end{figure}

WD~J2356$-$209 is not the only white dwarf to have a high Na/Ca abundance ratio.
A few objects with $\log\,{\rm Na/Ca}>0$ were identified by \cite{hollands2017cool}
and are shown in Figure \ref{fig:diffusion_evolution}.
However, because of the very noisy spectra of many of those objects,
the derived abundances are often highly uncertain. Using our own atmosphere model grid,
we performed a reanalysis of this sample and we found that, within the uncertainties,
most of those objects are compatible with a chondritic abundance ratio of $\log\,{\rm Na/Ca}=0$
(a detailed analysis will be presented in S.~Coutu et al., in preparation).
In fact, we found compelling evidence of a high Na/Ca abundance ratio only for
three objects (SDSS J095645.14+591240.6, SDSS J164939.23+223807.2 and SDSS J212312.20+001653.5).
Another potentially sodium-rich white dwarf, G77$-$50, was
analyzed by \cite{farihi2011magnetic}. They found a $\log {\rm Na/Ca}=0.7$ abundance
ratio, but this result should also be taken with a grain of salt as the magnetism of this
object complicates its analysis and the authors
explicitly noted that their sodium abundance measurement is uncertain.

\begin{figure}
    \includegraphics[width=\columnwidth]{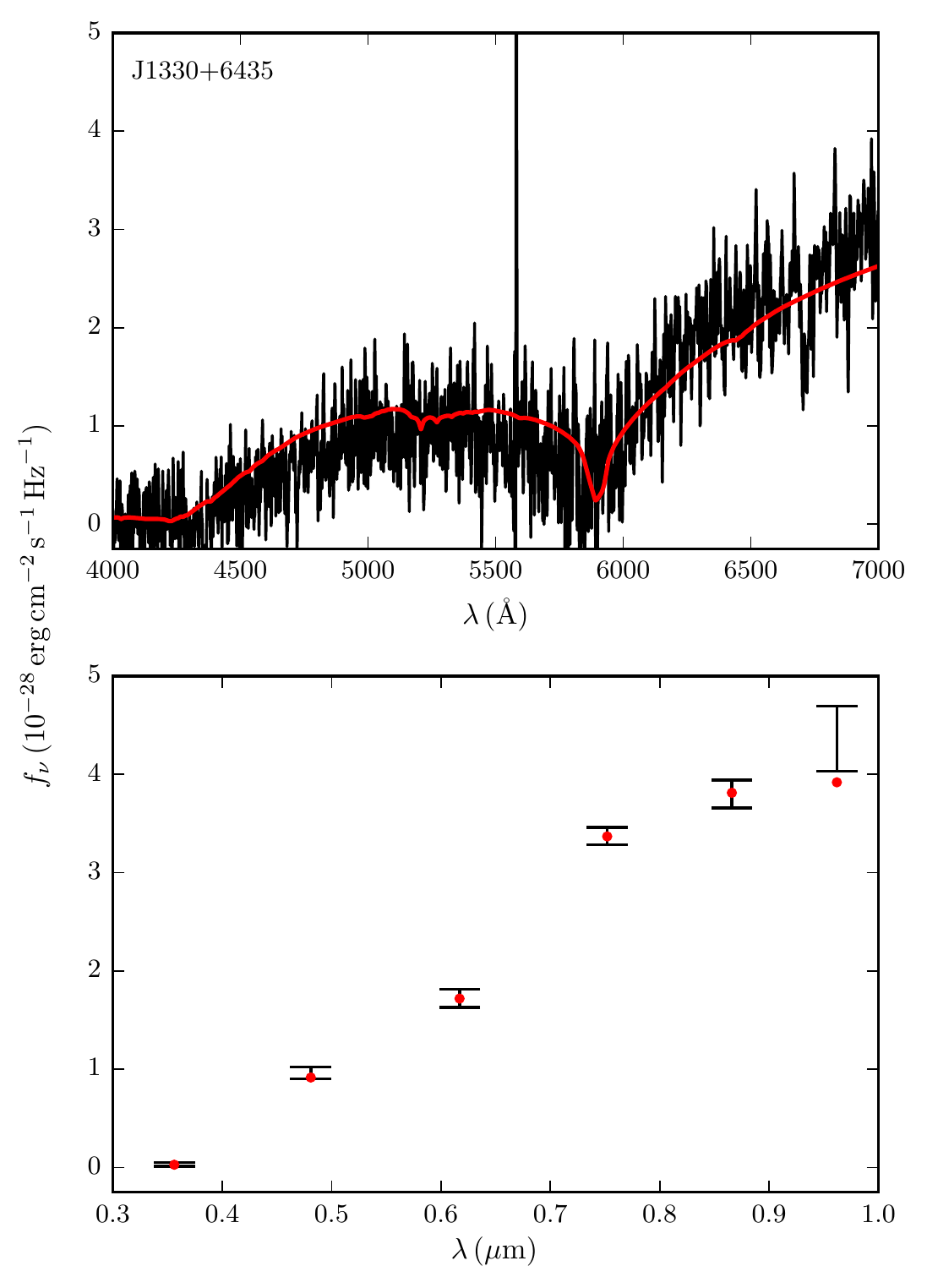}
    \caption{Our best solution for J1330+6435. The top panel shows our fit to the visible spectrum
      and the bottom panel displays our photometric fit to the SDSS $u$ band and Pan-STARRS $grizy$ bands.
      Note that the SDSS spectrum was smoothed for clarity.}
  \label{fig:J1330fit}
\end{figure}

Furthermore, \cite{harris2003initial} identified a faint ($g=21.4$) DZ star in the
Sloan Digital Sky Survey (SDSS) that has a spectrum that also appears to show a broad Na line.
As WD~J2356$-$209, the visible spectrum of SDSS J133001.13+643523.7 (J1330+6435) is
dominated by a strong and broad \ion{Na}{1} D doublet. Figure \ref{fig:J1330fit} shows our best solution
for this star and Table \ref{tab:1330} lists its derived atmospheric parameters. 
Note that a hydrogen-free atmosphere was assumed, since we do not have any infrared photometry to
fit the H/He ratio to an eventual infrared flux depletion.\footnote{In any case, the addition of a moderate
  amount of hydrogen (up to $\log\,{\rm H/He}=-1.5$) does not change our conclusions on the
  Na/Ca abundance ratio of this object.}
The SDSS spectrum being noisy, the Ca/He and Na/He abundance ratios are subject
to considerable uncertainties. Nonetheless, it is clear that the Na/Ca ratio of J1330+6435 is higher
than average, but well below the extreme ratio found for WD~J2356$-$209.
While there are some objects with a Na/Ca abundance ratio above the chondritic value,
WD~J2356$-$209 is the most extreme case ever encountered.

\begin{deluxetable}{cc}
  \tablecaption{J1330+6435 atmospheric parameters \label{tab:1330}}
  \tablehead{\colhead{\hspace{0.7cm}Parameter}\hspace{0.7cm} &
    \colhead{\hspace{0.7cm}Value}\hspace{0.7cm}}
  \startdata
  $T_{\rm eff}$       & $\phd \phn \phn 4310 \pm 190\,{\rm K}$\\
  $\log g$            & $\phn \phd 8.26 \pm 0.15$\\
  ${\rm H/He}$        & $\phn \phd 0$\\
  $\log\,{\rm Na/He}$ & $-8.5 \pm 0.3$\\
  $\log\,{\rm Ca/He}$ & $-8.8 \pm 0.3$
  \enddata
\end{deluxetable}

\subsection{Constraints on the accreted planetesimal}

The abundances of heavy elements in a DZ star are indicative of the nature of the
accreted parent body and can be used to infer its composition
\citep{zuckerman2007chemical,koester2011cool,dufour2012detailed,hollands2018cool,harrison2018polluted}.
Such studies have highlighted the existence of a great diversity of accreted bodies,
ranging from Kuiper-Belt-Object analogs \citep{xu2017chemical} to 
differentiated rocky bodies \citep{zuckerman2011aluminium,melis2017differentiated}
and water-bearing planetesimals \citep{farihi2013evidence,raddi2015likely}.
Can we relate the high Na/Ca ratio of WD~J2356$-$209 to any solar system counterpart?

\begin{figure}
    \includegraphics[width=\columnwidth]{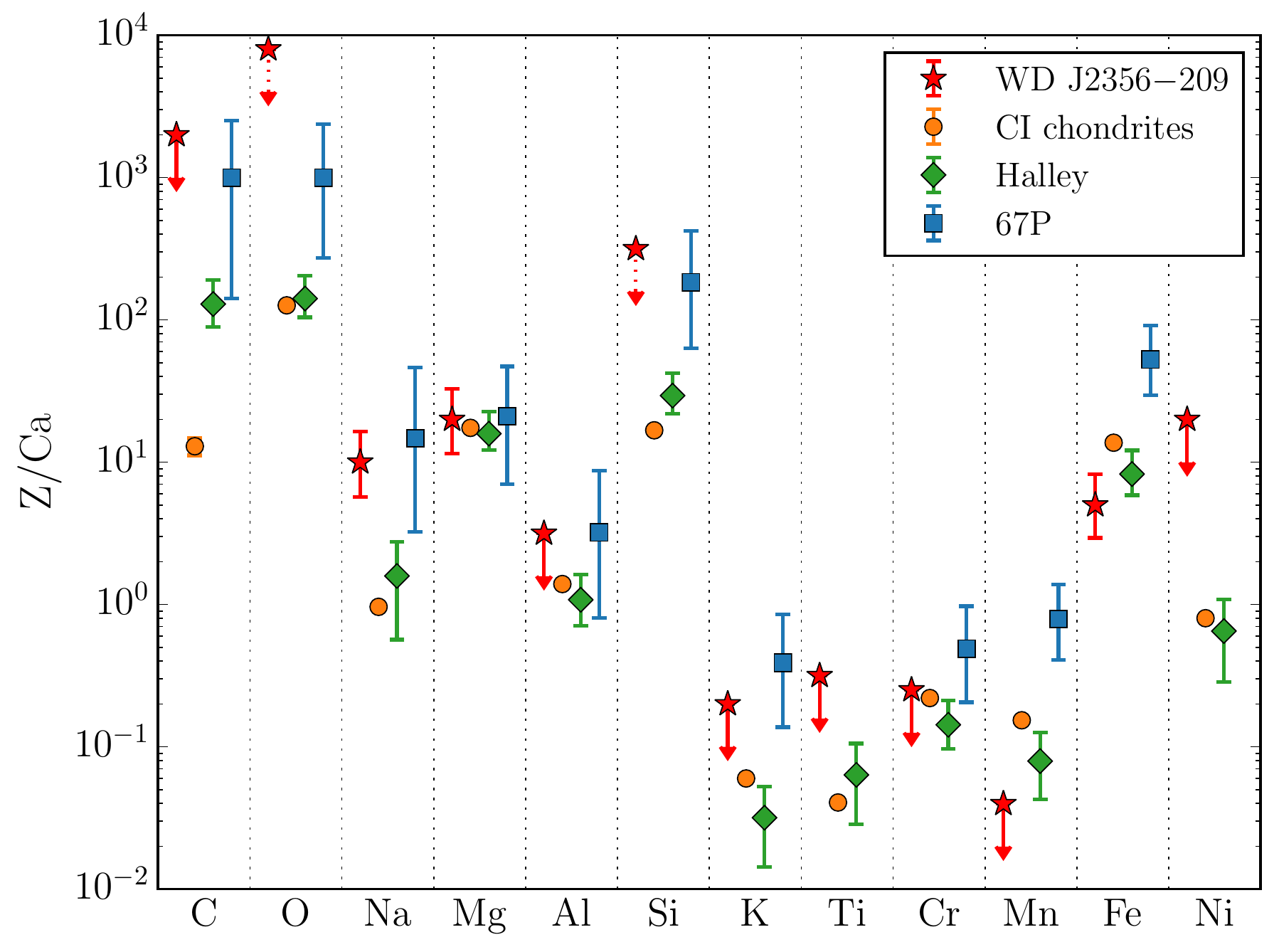}
    \caption{Elemental abundances by number relative to Ca. The data for WD~J2356$-$209 is
      from Table \ref{tab:2356}, chondritic abundances are taken from
      \cite{lodders2003solar} and the data for comets Halley and 67P are from
      \cite{jessberger1988aspects} and \cite{bardyn2017carbon}, respectively. 
      The upper limits represented by dotted arrows are those for which
      no firm constraint could be established. Note that the abundances
      for WD~J2356$-$209 correspond to the abundance ratios of the accreted planetesimal
      only if accretion is in its early phase (i.e., the effects of relative diffusion are 
      not included in this figure).
    }
  \label{fig:composition}
\end{figure}

Table \ref{tab:naca} lists the Na/Ca ratio of a few solar system objects. The only
measurements compatible with the composition of WD~J2356$-$209 are those obtained for the
dust of the comet 67P/Churyumov-Gerasimenko (67P). Those measurements were obtained
by the COmetary Secondary Ion Mass Analyzer (COSIMA, \citealt{kissel2007cosima}) 
onboard the \textit{Rosetta} spacecraft while it followed comet 67P at close distances
for two years. Additionally, with the exception of Mn and Fe, the abundances of the other
heavy elements in WD~J2356$-$209
are compatible with those measured in the dust of comet 67P (Figure \ref{fig:composition}).

\begin{deluxetable}{ccc}
  \tablecaption{Sodium abundances \label{tab:naca}}
  \tablehead{\colhead{Object} & \colhead{$\log\,{\rm Na/Ca}$\tablenotemark{a}} & \colhead{Reference}}
  \startdata
  WD~J2356$-$209 & $\phn \phd 1.0 \pm 0.2$ & This paper\\
  CI chondrites & $-0.017 \pm 0.002$ & \cite{lodders2003solar} \\
  Bulk Earth & $-0.64 \pm 0.04$ & \cite{wang2018elemental} \\
  Earth's mantle & $-0.74 \pm 0.04$ & \cite{wang2018elemental}\\
  Earth's crust & \phn 0.00 & \cite{crc99}\\
  Halley & $\phn \phd 0.2 \pm 0.3$ & \cite{jessberger1988aspects}\\
  67P & $\phn \phd 1.2 \pm 0.6$ & \cite{bardyn2017carbon}
  \enddata
  \tablenotetext{a}{Ratio of number densities}
\end{deluxetable}

That being said, the total mass accreted by WD J2356$-$209 is incompatible with the mass
of a comet. By estimating the mass of metals currently in the convection zone, we can
obtain a lower limit on the total accreted mass (it is a lower limit since diffusion
might already have evacuated heavy elements from the convection zone).
Using the envelope model code of \cite{rolland2018spectral}, we find that the fractional
mass of the convection zone is $\log q=-5.49$ for a $T_{\rm eff}=4000\,{\rm K}$ white dwarf with
$\log g=8$ and $\log\,{\rm H/He}=-1.5$.
With this value, we find that the total mass in the convection zone
of the four heavy elements clearly detected in WD~J2356$-$209
(Ca, Mg, Na and Fe) is $5 \times 10^{20}\,{\rm g}$. This value could significantly
underestimate the total mass of metals currently in the convection zone,
since it does not include C, O and Si,
three elements that are expected to be particularly abundant
(Figure \ref{fig:composition}). A higher estimate can be obtained by assuming
that the accreted planetesimal had a chondritic composition relative to Ca and computing its total
mass from the abundance of Ca detected in WD~J2356$-$209. Using the
values published in \cite{lodders2003solar}, this yields a mass of $\sim 10^{21}\,{\rm g}$.
This mass is significantly higher than the mass of comet 67P
\citep[$10^{16}\,{\rm g}$,][]{patzold2016homogeneous} and that of the most massive comets
of the solar system
\citep[Hale-Bopp has a mass of $\sim 10^{19}\,{\rm g}$,][]{weissman2007cometary}, suggesting
that a single comet cannot be the cause of the pollution of the atmosphere of WD~J2356$-$209.
As the lower limit on the accreted mass is $\sim 10^2$ times higher than the mass of the
largest comets, any scenario in which several comets were accreted can safely be rejected.

\subsection{Relative diffusion}

Alternatively, it is possible that the accreted planetesimal did not have an
unusual composition and the extreme Na/Ca abundance
ratio could simply be due to the different
diffusion timescales of Na and Ca. 
In fact, for a cool helium-rich white dwarf like WD~J2356$-$209, diffusion-accretion equilibrium
cannot be assumed \citep{koester2011cool,hollands2018cool}.
This follows from the fact that the diffusion
times are of the order of a few Myr
\citep{paquette1986diffusion,koester2009accretion,hollands2017cool}, while the typical
duration of an accretion episode is only of $10^4-10^6$ years
\citep{jura2008pollution,kilic2008near,melis2011accretion,girven2012constraints}.
It is therefore possible that WD~J2356$-$209 has stopped accreting and that the difference
between the diffusion timescales of Na and Ca has led to a composition that is not
representative of the accreted planetesimal.

Once the accretion episode has stopped, relative diffusion implies that a change
$\Delta \log\,{\rm Ca/He}$ will be accompanied by a variation
of the Na/Ca abundance ratio given by \citep[][Equation 3]{hollands2018cool},
\begin{equation}
  \frac{\Delta \log\,{\rm Na/Ca}}{\Delta \log\,{\rm Ca/He}} =
  \frac{\tau_{\rm Ca}}{\tau_{\rm Na}} -1.
  \label{eq:relative_diffusion}
\end{equation}
In WD~J2356$-$209, the diffusion timescale of Na is longer by a factor of $\approx 1.7$
than the diffusion timescale of Ca
($\log \tau_{\rm Ca} \approx 6.32$ and $\log \tau_{\rm Na} \approx 6.56$).\footnote{To estimate the diffusion timescales, we use the values given by
  \cite{hollands2017cool} for SDSS J163601.33+161907.1, as this star has similar
  atmospheric parameters to WD~J2356$-$209.}
From Equation \ref{eq:relative_diffusion}, it follows that every 1 dex decrease
in $\log\,{\rm Ca/He}$
is accompanied by a 0.4 increase in $\log\,{\rm Na/Ca}$. If we assume for a moment
that the accreted planetesimal had a chondritic composition with $\log\,{\rm Na/Ca}=0$,
it implies that the Ca abundance in the atmosphere of WD~J2356$-$209 was
$\log\,{\rm Ca/He}=-6.8$ when the accretion episode stopped, some $\approx 10\,{\rm Myr}$
ago \citep[][Equation 4]{kawka2016extreme}. This would raise our
lower limit on the total accreted mass to $\sim 10^{23}\,{\rm g}$, which corresponds
to the mass of a very large asteroid
\citep[Vesta has a mass of $2.6 \times 10^{23}\,{\rm g}$,][]{russell2012dawn}.
However, this scenario appears unlikely. Cool DZ stars are known to be much less
polluted than hotter objects
\citep{dufour2007spectral,koester2011cool,hollands2017cool}, possibly because the
largest planetesimals are gradually scattered away from the planetary system
\citep{veras2013simulations,veras2016full,hollands2018cool}.
In fact, according to the Montreal White Dwarf
Database \citep{dufour2016montreal}, no known DZ star cooler than
$T_{\rm eff}=5000\,{\rm K}$ has a calcium abundance higher than $\log\,{\rm Ca/He}=-9.3$.

More importantly, assuming that WD~J2356$-$209 accreted
a planetesimal with a chondritic composition might not be a realistic hypothesis.
Figure \ref{fig:diffusion_evolution} shows how the composition of WD~J2356$-$209 is expected
to have evolved given the diffusion timescales of \cite{hollands2017cool}. This
evolution rules out the possibility that the accreted planetesimal had a
chondrite-like composition. A more realistic
hypothesis would probably be to assume that the accreted planetesimal had a sodium abundance
ratio of $\log\,{\rm Na/Ca} \approx 0.6$, which corresponds to the region where
the chemical evolution track of WD~J2356$-$209 intersects the middle
of the distribution of objects plotted in Figure \ref{fig:diffusion_evolution}.
In that case, the Ca abundance in the atmosphere of WD~J2356$-$209 when the accretion
episode stopped
$\approx 4\,{\rm Myr}$ ago (i.e., $1.9 \tau_{\rm Ca}$ or $1.1 \tau_{\rm Na}$)
would have been a less extreme $\log\,{\rm Ca/He} \approx -8.3$ and
the lower limit on the accreted mass is decreased to $\sim 10^{22}\,{\rm g}$.
Assuming a typical asteroid density of $2\,{\rm g}\,{\rm cm}^{-3}$, the accreted
planetesimal had a radius of at least 100\,km.

\section{Conclusion}
\label{sec:conclusion}
We presented a detailed analysis of WD~J2356$-$209, a very cool DZ star whose
spectrum is dominated by a strong sodium feature. Thanks to our improved atmosphere models (Papers~I and II),
we find an excellent fit to the photometry and the visible spectrum, allowing the
first reliable atmospheric parameter determination of this object. We found that
WD~J2356$-$209 has a record sodium abundance with a number density ratio
$\log\,{\rm Na/Ca}=1.0 \pm 0.2$.

A possible explanation for this high sodium content is that WD~J2356$-$209 has accreted
a planetesimal with an abnormally high sodium abundance and a total mass
$\gtrsim 10^{21}\,{\rm g}$.
However, we were unable to identify a solar system analog to this hypothetical
planetesimal, since none of the examined candidates simultaneously matched the
constraints on the sodium abundance and the total mass.
Alternatively, the high sodium abundance in the atmosphere of WD~J2356$-$209 could
be explained by the slower diffusion of Na with respect to Ca. According
to this scenario, the accreted object would have had a less extreme composition
but a larger total mass ($\gtrsim 10^{22}\,{\rm g}$).

This paper concludes the observational validation of our improved
cool white dwarf model atmosphere code.
The excellent fit obtained for this extreme object is a clear demonstration of the
capacity of our models to properly take into account the nonideal effects that
prevail in the atmospheres of cool white dwarfs.
In the next paper of this series, we will use our improved models to analyze a large
sample of cool white dwarfs and revisit their spectral evolution.

\acknowledgments

We would like to thank Manuel Barranco for sharing with us ab initio potentials for the Ca$-$He interaction.
We are grateful to the anonymous reviewer for her or his valuable comments on our manuscript.
We also thank Siyi Xu for useful discussions regarding the composition of WD~J2356$-$209.

This work was supported in part by NSERC (Canada) and the Fund FRQNT (Qu\'ebec).
This work has made use of the Montreal White Dwarf Database \citep{dufour2016montreal}.

This work has made use of data from the European Space Agency (ESA) mission
{\it Gaia} (\url{https://www.cosmos.esa.int/gaia}), processed by the {\it Gaia}
Data Processing and Analysis Consortium (DPAC,
\url{https://www.cosmos.esa.int/web/gaia/dpac/consortium}). Funding for the DPAC
has been provided by national institutions, in particular the institutions
participating in the {\it Gaia} Multilateral Agreement.

The Pan-STARRS1 Surveys (PS1) and the PS1 public science archive have been made possible through contributions by the Institute for Astronomy, the University of Hawaii, the Pan-STARRS Project Office, the Max-Planck Society and its participating institutes, the Max Planck Institute for Astronomy, Heidelberg and the Max Planck Institute for Extraterrestrial Physics, Garching, The Johns Hopkins University, Durham University, the University of Edinburgh, the Queen's University Belfast, the Harvard-Smithsonian Center for Astrophysics, the Las Cumbres Observatory Global Telescope Network Incorporated, the National Central University of Taiwan, the Space Telescope Science Institute, the National Aeronautics and Space Administration under Grant No. NNX08AR22G issued through the Planetary Science Division of the NASA Science Mission Directorate, the National Science Foundation Grant No. AST-1238877, the University of Maryland, Eotvos Lorand University (ELTE), the Los Alamos National Laboratory, and the Gordon and Betty Moore Foundation.

This publication makes use of data products from the Wide-field Infrared Survey Explorer, which is a joint project of the University of California, Los Angeles, and the Jet Propulsion Laboratory/California Institute of Technology, funded by the National Aeronautics and Space Administration.

\bibliographystyle{aasjournal}
\bibliography{references}

\end{document}